\documentclass{moriond}
\newcommand{\Photo}{}
\usepackage{float}
\usepackage{multirow}
\usepackage{geometry}
\usepackage{amsmath}
\usepackage{array}

\numberwithin{figure}{section}
\numberwithin{table}{section}
\numberwithin{equation}{section}

\def\be{\begin{equation}}
\def\ee{\end{equation}}
\def\bea{\begin{eqnarray}}
\def\eea{\end{eqnarray}}

\title{Search for new physics in $B\to K\pi\pi\gamma$
with Belle II data}

\author{ Sahil Saha }
\author{on behalf of the Belle II collaboration}
\address{IPHC Strasbourg, 23 Rue du Loess, 67200 Strasbourg, France }
\begin{document}

\vspace*{4cm}

\maketitle

\abstracts{
The measurement of the time-dependent \emph{CP}-asymmetry of \(B^0 \to K_S^0 \pi^+ \pi^- \gamma\) decays is sensitive to contributions from physics beyond the Standard Model. To translate this measurement into a constraint on new physics, it is necessary to distinguish between decay modes that contribute to the final state via a kaonic resonance that is a \emph{CP} eigenstate, and those that proceed through non-\emph{CP} eigenstates. This requires an amplitude analysis of the \(B \to K_{res} \gamma \to K \pi^+ \pi^- \gamma \) decay, that is discussed in this article.}

\section{Introduction} \label{Introduction}

\begin{figure}[H]
    \centering
    \includegraphics[width=0.275\linewidth]{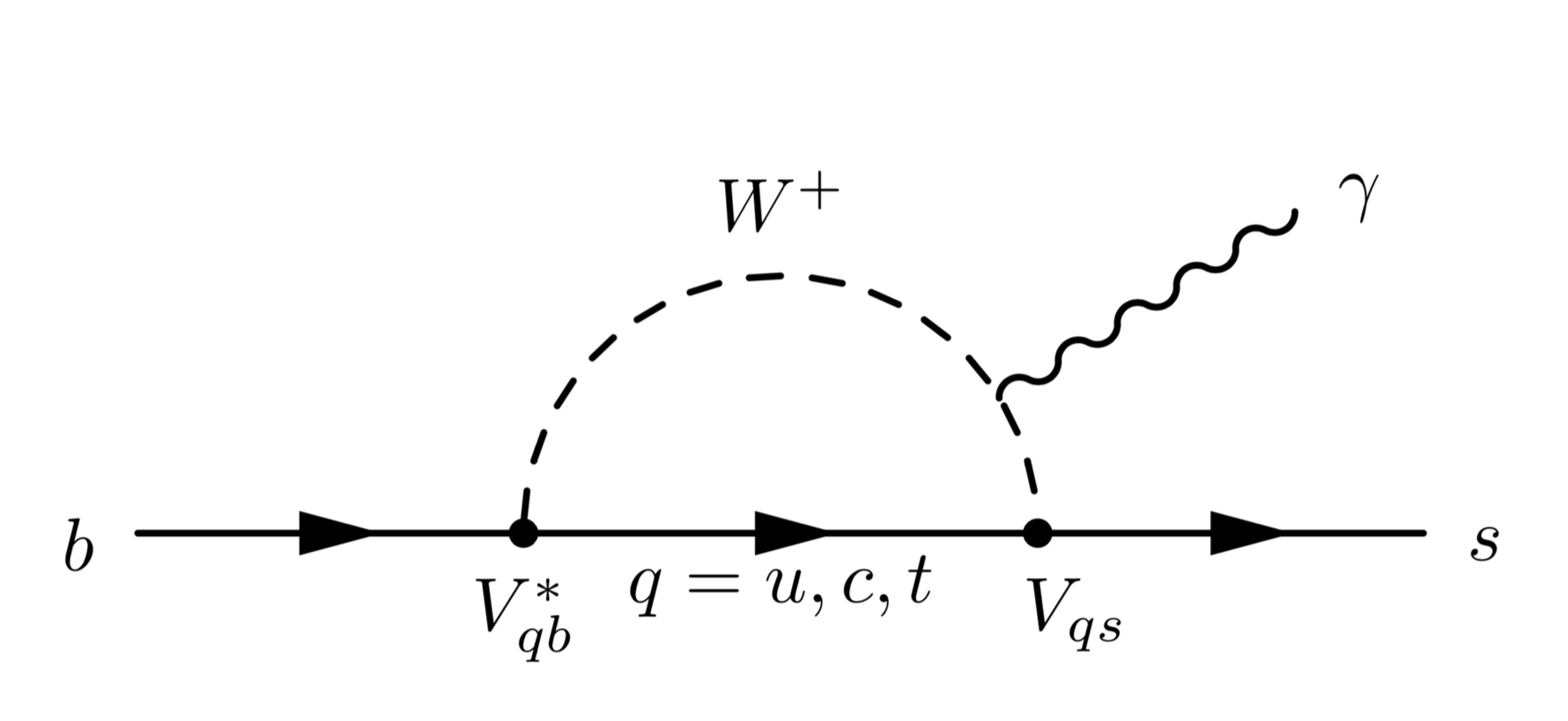}
    \caption{Feynman diagram for the \(b\to s \gamma\) transition.}
    \label{Fig:penguin}
\end{figure}

Radiative \(b\) decays, such as \(b\to s \gamma\) transitions, proceed through flavor-changing neutral currents (FCNC) forbidden at tree level in the Standard Model (SM). The FCNC process can only occur at first order through the penguin diagram displayed in Fig. \ref{Fig:penguin}. This process is suppressed in the SM due to the Glashow–Iliopoulos–Maiani (GIM) mechanism.  The possible contribution of a new heavy particle in the electroweak
loop would cause the process to deviate from the SM prediction.

The measurement of the \emph{CP}-asymmetry\cite{Akar_2019} of \(B^0 \to K_S^0 \rho^0(\pi^+ \pi^-) \gamma \) as a function of time between the \(B\) and \(\bar B\) decays - \(t\) and is expressed as:
\begin{equation} \label{tdcpasymm}
    \mathcal A_{{CP}}(t) = \frac{\Gamma(\bar B^0(t) \to f_{{CP}}\; \gamma)-\Gamma( B^0(t) \to f_{{CP}}\; \gamma)}{\Gamma( \bar B^0(t) \to f_{{CP}}\; \gamma)+\Gamma( B^0(t) \to f_{{CP}}\; \gamma)} =  \mathcal S_{{CP}} \sin (\Delta m t) - \mathcal C_{{CP}} \cos (\Delta m t).
\end{equation}
Here, \(f_{CP}\) is the \emph{CP}-eigenstate \( K_S^0 \rho^0 \), \(\Gamma\) is the decay rate and \( \Delta m\) is the mass difference of the neutral \(B\) meson mass eigenstates. $\mathcal{S}_{CP}$ and $\mathcal{C}_{CP}$ are the mixing-induced and direct \emph{CP} asymmetries respectively.
The presence of the non-\emph{CP} eigenstates going to the same final state dilutes the \emph{CP} contributions to the asymmetry. Therefore, an effective \emph{CP} asymmetry - \(\mathcal S_{B^0 \to K_S^0 \pi^+ \pi^- \gamma}\) is defined and it is related with \( \mathcal{S}_{CP}\) with a dilution factor \cite{Akar_2019} $\mathcal{D}$: 
\begin{equation} \label{tdcpseff}
    \mathcal S_{B^0 \to K_S^0 \pi^+ \pi^- \gamma} = \mathcal D \cdot \mathcal S_{CP}.
\end{equation}
The effective \emph{CP} parameter \(\mathcal{S}_{B^0 \to K_S^0 \pi^+ \pi^- \gamma} = -0.29 \pm 0.11 \pm 0.05\) was measured using Belle and Belle II data \cite{2026}.
\par 
In order to constrain new physics without knowing explicitly its field content, we can express the \(b \to s \gamma\) process in terms of an effective field theory, where the effective Hamiltonian for the process would be \cite{Paul_2017}:
\begin{equation} \label{hamiltonian}
    \mathcal H_{eff} = - \frac{4 G_F}{\sqrt{2}}V_{tb}V_{ts^*}\left(\sum_{i=1}^8C_iQ_i +C'_iQ'_i\right).
\end{equation}
Here, \(G_F\) is the Fermi constant, \(C_i/C_i'\) are the short-distance Wilson 
coefficients, and \(Q_i/Q_i'\) are the local long-distance operators. 
The operators \(Q_7\) and \(Q_7'\) describe left and right-handed chiral 
structures of \(b \to s \gamma\). In the SM, \(C_7'\) arises from external 
quark mass insertions, giving \(C_7'/C_7 \sim m_s/m_b \approx 0.02\). 
Thus, \(b \to s \gamma\) produces a predominantly left-handed photon 
(right-handed for \(\bar{b} \to \bar{s} \gamma\)). The \emph{CP}-asymmetry parameter \(\mathcal{S}_{CP}\) can be written as a function of the Wilson coefficients \cite{Akar_2019} -
\begin{equation} \label{S_C7}
    \mathcal{S}_{CP}  = -\frac{2\text{Im}(C_7C_7')}{|C_7|^2+|C_7'|^2}.
\end{equation} 
The dilution factor \(\mathcal D\) is a function of the decay amplitude \(A(p1,p2,p3)\).
\begin{equation} \label{dilfac}
    \mathcal{D} = -\frac{\int \text{Re} \left( A(p1,p2,p3) \cdot A(p2,p1,p3)\right) dp}{\int |A(p1,p2,p3)|^2dp},
\end{equation}
where, \(A(p1,p2,p3) = A\left(B^0 \to K_S^0 (p_3) \pi^+ (p_1) \pi^- (p_2) \gamma\right)\).This dilution factor can be measured from a complete amplitude analysis of the \(B^0 \to K^0_S \pi^+ \pi^- \gamma\) decay. In practice, we perform the amplitude analysis of the charged isospin partner \(B^+ \to K^+ \pi^+ \pi^- \gamma \) decay (the charge-conjugated decay of \(B^-\) is also included assuming no direct CP violation), as it is statistically favourable.

\section{Decay model} \label{Decay Model}
\begin{figure}[H]
    \centering
    \begin{minipage}{0.45\textwidth}
        \begin{center}
            \begin{tabular}{|c|l|}
            \hline
            \rule{0pt}{1.3em} $\mathbf{J^{\mathbf{P}}}$ &  \hspace{0.4in} Decay Mode \hfill \\
            \hline \hline
            \multirow{5}{*}{$1^+$} 
            \rule{0pt}{1em}
            &  $K_1(1270)^+ \to K^*(892)^0 \pi^+ $ \\
            & $K_1(1270)^+ \to K^+\rho(770)^0 $ \\ 
            & $K_1(1270)^+ \to K^+\omega(782)^0 $ \\
            & $K_1(1270)^+ \to K^*(1430)^0 \pi^+ $ \\
            &  $K_1(1400)^+ \to K^*(892)^0 \pi^+ $ \\
            &  $K_1(1400)^+ \to K^+\rho(770)^0 $ \\
            \hline
            \multirow{3}{*}{$1^-$} 
            &  $K^*(1410)^+ \to K^*(892)^0 \pi^+$ \\
            &  $K^*(1410)^+ \to K^+\rho(770)^0$ \\
            & $K^*(1680)^+ \to K^*(892)^0 \pi^+$ \\ 
            & $K^*(1680)^+ \to  K^+\rho(770)^0$ \\
            \hline
            \multirow{3}{*}{$2^+$} 
            &  $K_2^*(1430)^+ \to K^*(892)^0 \pi^+$ \\
            & $K_2^*(1430)^+ \to  K^+\rho(770)^0$ \\ 
            & $K_2^*(1430)^+ \to K^+\omega(782)^0 $ \\
            \hline
            \multirow{4}{*}{$2^-$} 
            & $K_2(1770)^+ \to K^*(892)^0 \pi^+ $ \\
            & $K_2(1770)^+ \to K^+\rho(770)^0 $ \\
            & $K_2(1770)^+ \to K_2^*(1430)^0 \pi^+ $ \\ 
            & $K_2(1770)^+ \to K^+ f_2(1270)^0 $ \\
            \hline
        \end{tabular}
        \caption{Table of intermediate kaonic resonances in the \(B^+ \to K^+ \pi^+ \pi^- \gamma\) decay.}
        \end{center}
        \label{tab:decays}
    \end{minipage}%
    \hfill
    \begin{minipage}{0.45\textwidth}
        \centering
        \includegraphics[width=\linewidth]{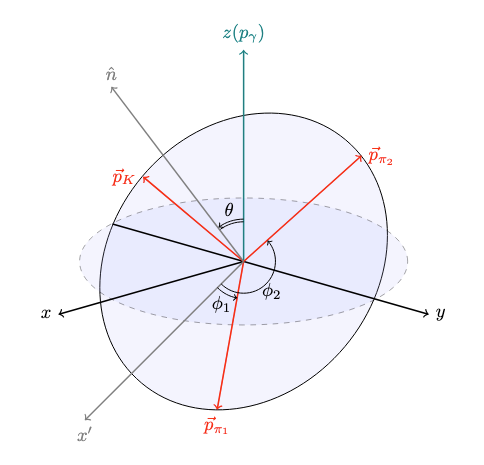}
        \caption{Kinematics of the $B^+ \to K^+ \pi^+ \pi^- \gamma$ decay.}
        \label{fig:kinematics}
    \end{minipage}
\end{figure}

In this section, the model describing the \(B^+ \to K^+ \pi^+ \pi^- \gamma \) decay is outlined. This decay proceeds through three intermediate steps. Firstly, the \(B^+ \to K_{res}^+ \gamma\) step. Each possible \(K_{res}^+\) can be of the spin parity \(J^P = 1^\pm,2^\pm\) and interfere among each other. Then the \(K_{res}^+\) decays into \(\rho^0 K^+\), \( K^{*0}\pi^+\) or \( (K\pi)_S^0 \pi^+\) which finally decay to the \( K^+ \pi^+ \pi^-\) state. The list of intermediate paths to the $K^+ \pi^+ \pi^-$ final state are listed in Table \ref{tab:decays}.

To account for the possible quantum interferences between all decay modes, a coherent sum of the amplitudes of each individual decay mode is necessary. The final probability is then:
\begin{equation} \label{probability}
    P_{total} = |\sum_i c_i A_i|^2,
\end{equation}

where \(c_i\) and \(A_i\) are respectively the complex weights and amplitudes of each decay path. Thus, the amplitude for each intermediate path can be expressed as a product of three consecutive decays:
\begin{equation} \label{ampexpression}
    A_i = (S_B \cdot B_{L_B})\cdot (T_{K_{res}} \cdot S_{K_{res}}\cdot B_{L_{K_{res}}}) \cdot (T_X \cdot {S_X}\cdot B_{L_X}),
\end{equation}

where \(S\) is the spin-dependent part, dictating the angular distribution of the decay, \(T\) denotes the lineshapes of the \(K_{res}\) and of \(X\) which can be \(\rho^0, K^{*0} \text{ or } (K\pi)_S^0\). It must be noted that in our model, the contribution of \((K\pi)_S^0\) is included in the decay of \(K_1(1270)^+ \to K^*(1430)^0 \pi^+\).
The spin-dependent part of the amplitudes is computed with the covariant tensor formalism for the decay of $K_{res} \to K^+ \pi^+ \pi^-$, and the \(B\) decay expression includes radiative corrections\cite{hebinger:tel-01915429}.
To estimate the contributions of each possible intermediate path from a fit to Belle II data, we describe the amplitude in terms of five observables: three invariant masses \(m_{K^+\pi^+\pi^-}, m_{K^+\pi^-}\) and \(m_{K^+\pi^+\pi^-}\), and two angles \(\cos\theta\) and \(\phi_1\). The kinematics of the system and the definition of the cartesian frame is shown in Fig. \ref{fig:kinematics}. The angles are defined in the \(K_{res}\)-rest frame with the photon along the \(z\)-axis while the \(y\)-axis is oriented towards \(\vec p_{\pi^+} \times \vec p_{\pi^-}\). \(\cos\theta\) is the angle between the normal \(\hat n\) to the \(K^+\pi^+\pi^-\) plane and the photon. (\(\phi_{1/2}\)) are the angles of the pions with respect to \(\hat x' = \hat n \times \hat y\). To describe the amplitude, either \(\phi_{1/2}\) or their mean \(\phi\) can be used. 

\section{Preliminary developments of the amplitude analysis} \label{Amplitude Analysis}
The complex weights described in Section \ref{Decay Model} will be obtained by fitting our decay model to \(B^+ \to K^+ \pi^+ \pi^- \gamma \) candidates reconstructed and selected from the Belle II dataset, with the same criteria as in the published measurement of the time-dependent CP asymmetry of $B^0 \to K^0_S \pi^+ \pi^- \gamma$ decays \cite{2026}.
Belle II data is collected from $e^+e^-$ collisions produced by SuperKEKB at a center-of-mass energy of $10.58$ GeV in order to produce the \(\Upsilon(4S)\) resonance decaying into \(B^+B^-\) pairs. \par
\begin{figure}[H]
    \centering
    \includegraphics[width=0.35\linewidth]{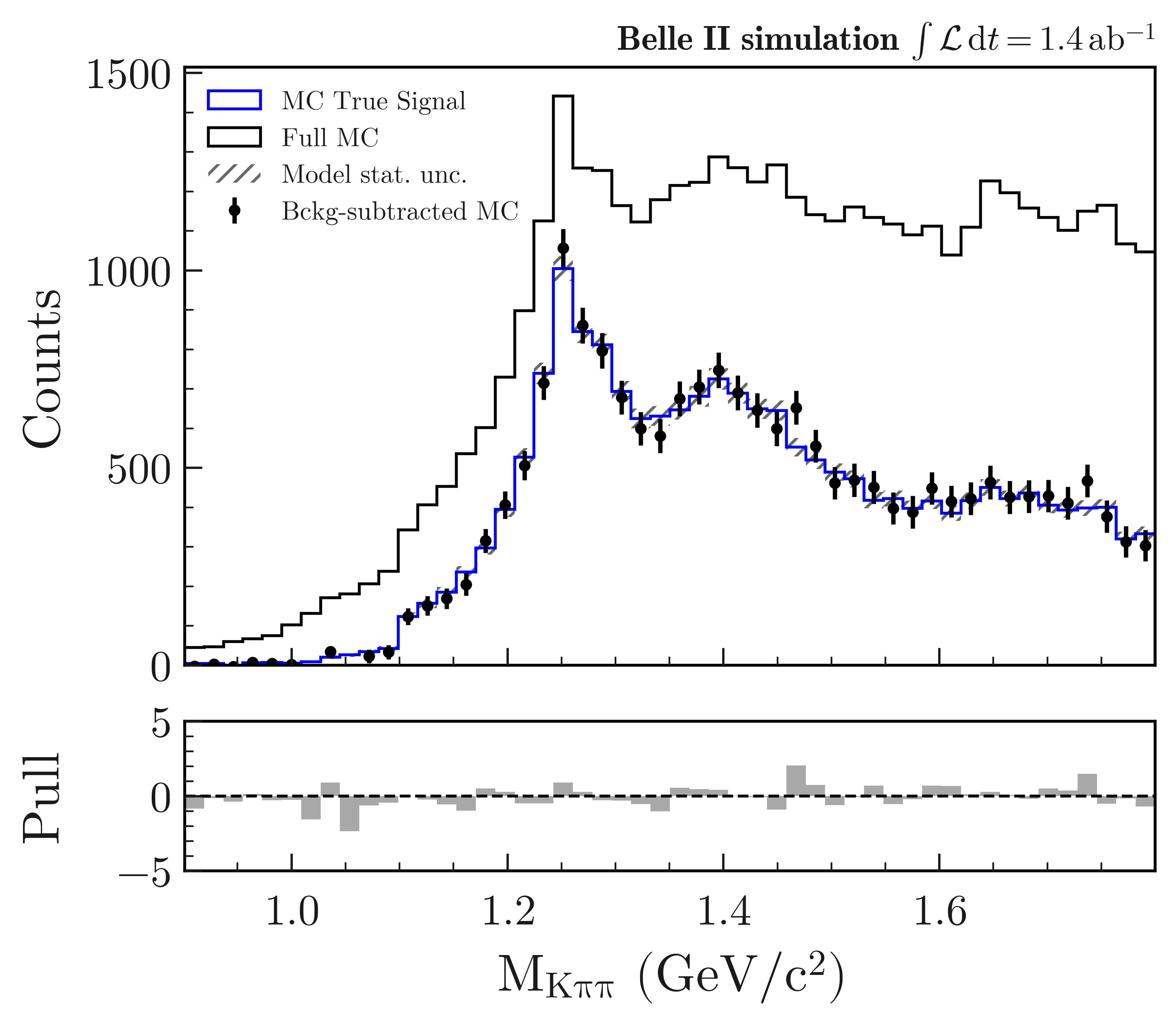}
    \caption{Simulated $m_{K^+\pi^+\pi^-}$ distribution after background subtraction with the sPlot method.}
    \label{fig:splot}
\end{figure}

Background-subtracted distributions of the five fit variables are obtained using the sPlot method \cite{Pivk_2005}. The discriminating variable between signal and background is \(\Delta E = \frac{2p_Bp_0 - s}{2\sqrt{s}}\), where \(p_B\) is the momentum of the \(B\) meson candidate, \(p_0 = p_{e^+} + p_{e^-}\) is the sum of the momenta of the beam particles and \(s\) is the beam-energy. As an illustration, Fig. \ref{fig:splot} shows the simulated \(m_{K^+\pi^+\pi^-}\) background-subtracted distribution corresponding to $1.4$ ab$^{-1}$ of Belle II data. \par 
The standard Monte-Carlo generator EvtGen\cite{Ryd:2005zz} does not account for interferences between the different $K_{res}$ modes and lacks the proper angular structure of the decay. Thus, we use AmpGen to generate the $B^+ \to K^+ \pi^+ \pi^- \gamma$ decay and process it through through the global Belle II simulation and reconstruction software.
\par
\begin{figure}[H]
    \centering
    \includegraphics[width=0.85\linewidth]{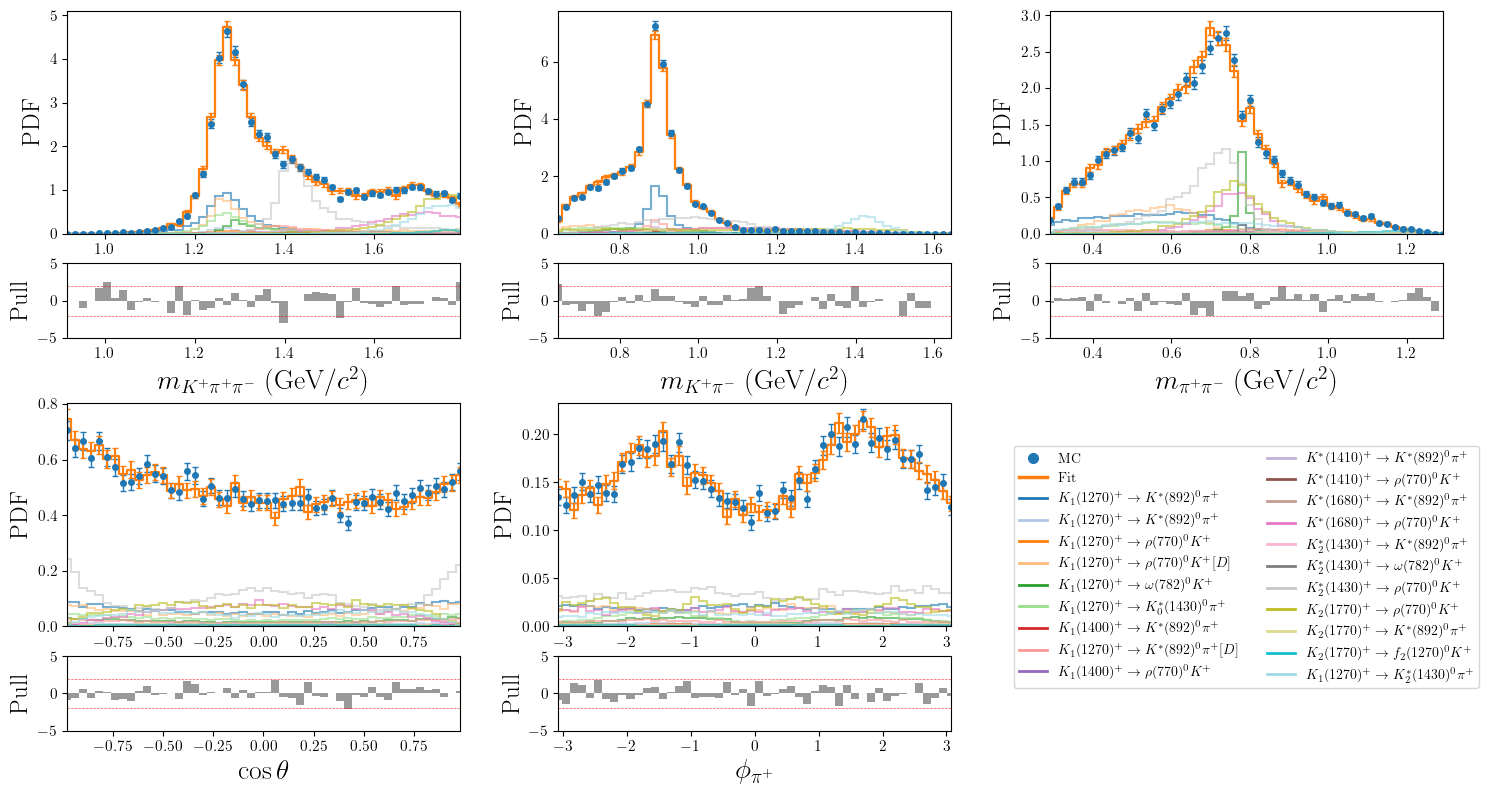}
    \caption{Fit to generator level simulated data for amplitude weights performed on a combination of all decay pathways from Table \ref{tab:decays}.}
    \label{fig:examplefit}
\end{figure}

Fig. \ref{fig:examplefit} illustrates the amplitude fit result. The fit function is the decay model discussed in section 2 taking into account all possible intermediate path in this example. The magnitudes and phases of each intermediate mode are left as free parameters. Pole masses and widths of each resonance are fixed to the PDG average values. The result of the fit is all the amplitudes and phases of each intermediate mode, that will allow to compute the total amplitude and further the dilution factor $\mathcal{D}$ using Eq. \ref{dilfac}.

\section{Conclusions and prospects}

The amplitude analysis formalism of \(B^+\to K_{res}^+\gamma \to K^+ \pi^+ \pi^- \gamma \) decays in Belle II was developed for the first time to describe all possible intermediate states. The next step will be to test the fit robustness and define the unbiased fit strategy, before performing the analysis of the Belle II data registered so far.

\section{Acknowledgements}

This work from the Interdisciplinary Thematic Institute QMat, part of the ITI 2021-2028 program at the University of Strasbourg, CNRS, and Inserm, was supported by IdEx Unistra (ANR 10 IDEX 0002), the SFRI STRAT’US project (ANR 20 SFRI 0012), and EUR QMAT (ANR-17-EURE-0024) under the French Investments for the Future Program.

\section*{References}
\bibliography{moriond}


\end{document}